\documentclass[twocolumn,prb,showpacs,10pt]{revtex4}
\usepackage{psfrag}
\usepackage{graphicx}
\usepackage{amsmath}
\usepackage{amssymb}
\usepackage{subfig}

\usepackage{color}
\usepackage{multirow}
\begin{document}
\title{Density functional theory based calculations of the transfer integral in a redox-active single molecule junction}
\author{Georg Kastlunger and Robert Stadler} 
\affiliation{Department of Physical Chemistry, University of Vienna, Sensengasse 8/7, A-1090 Vienna, Austria \\ 
Email: robert.stadler@univie.ac.at}

\date{\today}

\begin{abstract}
There are various quantum chemical approaches for an ab initio description of transfer integrals within the framework of Marcus theory in the context of electron transfer reactions. In our article we aim to calculate transfer integrals in redox-active single molecule junctions, where we focus on the coherent tunnelling limit with the metal leads taking the position of donor and acceptor and the molecule acting as a transport mediating bridge. This setup allows us to derive a conductance, which can be directly compared with recent results from a non-equilibrium Green's function approach. Compared with purely molecular systems we face additional challenges due to the metallic nature of the leads, which rules out some of the common techniques, and due to their periodicity, which requires {\bf k} space integration. We present three different methods, all based on density functional theory, for calculating the transfer integral under these constraints, which we benchmark on molecular test systems from the relevant literature. We also discuss manybody effects and apply all three techniques to a junction with a Ruthenium complex in different oxidation states.
\end{abstract}
\maketitle

\begin{section}{Introduction}\label{sec:intro}

In ultrahigh vacuum and at very low temperatures the electron transport problem in single-molecule junctions is nowadays straightforwardly accessible to a computational treatment with a nonequilibrium Green's function (NEGF) approach~\cite{keldysh} in combination with a density functional theory (DFT) based description of the electronic structure of the separate and combined components of the junction, namely the leads and the scattering region~\cite{atk}$^-$~\cite{kristian}. The theoretical modelling of experiments with an electrochemical scanning tunnelling microscope (STM)~\cite{Tim1,Tim2,Nichols1,Nichols2} is more challenging, because here depending on the setup as well as structural details of the system, two competing electron transport mechanisms have to be considered, namely electron hopping which is a thermally induced multiple step process and coherent tunnelling which is the standard one-step phenomenon known from benchmark molecules relatively strongly coupled to metallic electrodes at temperatures close to 0 K. In both cases an atomistic description of the process under electrochemical conditions provides a formidable challenge for a DFT based theory. For the former, the difficulty lies in a simplified and compact but nevertheless sufficiently accurate description of the nuclear vibrations of the molecule and solvent which drive the electron flow. For the latter it becomes necessary to adjust the oxidation state of the redox active center in the scattering region and therefore deal with the issue of charge localization in a multi-component system, which we addressed in a recent publication~\cite{first} where we also established a connection to our earlier work on electronegativity theory, Fermi level alignment and partial charge distributions within a single-molecule junction~\cite{robert-16,robert-13,robert-5}.

In our current article we focus on calculating the transfer integral~\cite{newton} in a single molecule junction, which is a key ingredient in the semi-classical Marcus theory often used for the description of electron hopping in purely molecular systems. This is a first step in treating hopping and coherent tunnelling on the same theoretical level, which enables a direct comparison of the coherent tunnelling conductance calculated from Marcus theory with that obtained from a NEGF approach and lays the ground for a description of electron hopping in our future work, where the reorganization energy and driving force will also have to be considered. For the quantum chemical description of the transfer integral there are two types of commonly used frameworks: 1) those that look for the minimum adiabatic state splitting, which is estimated either through Koopman's theorem~\cite{jordan} or by tuning energy differences with external perturbations~\cite{newton}; 2) those that depend on defining the diabatic states, such as the Mulliken-Hush method~\cite{hush} and its generalisation~\cite{newton1}, the block diagonalization method~\cite{newton2} and the fragment charge difference method~\cite{voityuk}. Because Slater determinant based techniques are rather unsuitable for the description of metallic systems (this holds also for parametrized approaches, which are very popular for the description of electron transfer processes in organic solids~\cite{pourtois}) , methods for both i) and ii) have also been developed within the framework of DFT more recently. 

We use three of these methods in our article. Within category 1 we employ an energy gap approach~\cite{newton,voityuk1}, where we define the splitting of adiabatic states in a single particle version as the energy difference between suitably selected Kohn Sham (KS) orbitals of our system and in a many body version, where we use the generalized $\Delta$SCF technique~\cite{Schiotz1,Schiotz2,Pawel1,Pawel2} for localizing a charge in particular orbitals, thereby including also electronic relaxation effects in our estimation of the energy splitting of adiabatic states at the transition point. For exploiting category 2, namely the derivation of transfer integrals via the definition of diabatic states, we use two fundamentally different approaches. One is based on calculating explicitly the coefficients for the expansion of adiabatic into diabatic states~\cite{migliore1,migliore2,migliore3}, where we introduce again a single particle and a many body version. The other is Larsson's formula for the estimation of an effective coupling~\cite{voityuk,larsson} , which is a multistate approach, where the contributions of all bridging molecular orbitals (MOs) are summed up. This last approach we use only with KS orbitals, because we could not find a meaningful many body implementation within DFT.

The paper is organized as follows: In the next section we introduce the three different methods for calculating the transfer integral in our article by applying them to a 3x3 tight binding (TB) Hamiltonian, which was also previously used by others for demonstration purposes~\cite{voityuk2}. In Section~\ref{sec:molecules} we describe how we combine the three techniques with DFT calculations, which we perform by using the GPAW code~\cite{GPAW1,GPAW2} and benchmark our approaches by comparing our results for molecular systems previously studied by other groups, namely for inter-molecular hole transport in a diethylene dimer~\cite{bredas} and for intra-molecular electron transport in a tetrathiafulvalene-diquinone anion (Q-TTF-Q$^-$)~\cite{vanvoorhis}. In section~\ref{sec:junction} we employ all techniques for the evaluation of the transfer integral in a single molecule junction with two gold electrodes connected by a Ru-complex with transport mediating MOs around the Fermi level. Here we study the influence of the representation of the leads, starting with four-atom gold clusters and ending up with a periodic slab description of the surface, where k point integration becomes an important issue~\cite{kpoints}. By using Nitzan's equations~\cite{nitzan1,nitzan2} we relate the conductance to the transfer integral in Marcus theory, where for the coherent tunnelling limit at low bias the reorganization energy and driving force can be disregarded, and we get reasonable quantitative agreement with our previous results, where we used a NEGF-DFT technique to calculate transmission functions for the same system~\cite{first}.

\end{section}

\begin{section}{Three methods for the evaluation of the transfer integral}\label{sec:methods}

In this section we introduce all three methods used in this article for calculating transfer integrals by applying them to the diabatic Hamiltonian $\bf{H}$ of order 3x3,
\begin{eqnarray}
\bf{H} =   \left(
\begin{array}{ccc}
\varepsilon_D & V_{DB} & V_{DA} \\
V_{DB} & \varepsilon_{B} & V_{BA} \\
V_{DA} & V_{BA} & \varepsilon_{A} 
\end{array}\right), \label{hamiltonian}
\end{eqnarray}
where $\varepsilon_D$, $\varepsilon_B$ and $\varepsilon_A$ are the onsite energies of a donor, a bridge and an acceptor state, respectively, and V$_{DB}$, V$_{DA}$ and V$_{BA}$ the respective electronic couplings between them. When we now specify the parameters in this Hamiltonian with $\varepsilon_A$=$\varepsilon_D$= 0, $\varepsilon_B$=1, V$_{DB}$=V$_{BA}$=-0.1 and V$_{DA}$=-0.01, which is representative for typical molecular donor/bridge/acceptor systems and identical with the setup studied in Ref.~\cite{voityuk2} a diagonalization of $\bf{H}$ results in the adiabatic states,
\begin{eqnarray}
\begin{array}{lllll}
\psi_1 & = 0.701 \phi_D & + 0.136 \phi_B & + 0.701 \phi_A,  &   \varepsilon_1 = -0.029 \\
\psi_2 & = 0.701 \phi_D & - 0.701 \phi_A, &  &   \varepsilon_2 = 0.010 \\
\psi_3 & = 0.096 \phi_D & - 0.991 \phi_B & + 0.096 \phi_A,  &   \varepsilon_3 = 1.019 . \label{adiabatic}
\end{array}
\end{eqnarray}

If we use the energy gap method~\cite{newton,voityuk1} for evaluating the transfer integral, we obtain the expression
\begin{equation}
H_{DA}^{gap} = (\varepsilon_2-\varepsilon_1)/2, \label{gap}
\end{equation}
where it is important to note that the eigenenergies of the adiabatic states $\psi_1$ and $\psi_2$ have been selected in this definition because of their high amplitudes on the donor and acceptor states, while the third adiabatic state $\psi_3$, which is mostly localised on the bridge state can be disregarded. In praxis, as we will also discuss in the following sections, there are always two distinct adiabatic states which can be used for forming the energy difference in Eqn.~\ref{gap} even for larger Hamiltonians as long as the donor and the acceptor are characterized by a single state on each side~\cite{voityuk1}.

Another definition of the transfer integral can be obtained by Larsson's formula for the derivation of an effective coupling~\cite{voityuk,larsson}
\begin{equation}
H_{DA}^{effect} =  V_{DA} - \Sigma_{i=1}^N \frac{V_{Di} V_{iA}}{\varepsilon_{A,D}-\varepsilon_i}, \label{effective}
\end{equation}
where the direct coupling between donor and acceptor V$_{DA}$ as well as the contributions from all N bridge states in an arbitrary system are added up explicitly and N=1 for the 3x3 Hamiltonian in Eqn.~\ref{hamiltonian}.

For the third technique we employ for calculating the transfer integral, we follow the work of Migliore~\cite{migliore1,migliore2,migliore3} and use the amplitudes on the donor and acceptor sites, i.e. the expansion coefficients a$_{D,1}$ and a$_{A,1}$, respectively, of the wavefunction for the adiabatic state with the lowest energy (the ground state $\psi_1$) in Eqn.~\ref{hamiltonian} to formulate
\begin{equation}
H_{DA}^{coeff} = \frac{a_{D,1} a_{A,1}}{a_{D,1}^2 - a_{A,1}^2} (\varepsilon_{A} - \varepsilon_{D}) . \label{direct}
\end{equation}
Since the diabatic states in Eqn.~\ref{hamiltonian} are orthogonal to each other by definition, we do not need to apply the orthogonalisation procedure detailed in Refs.~\cite{migliore1}-~\cite{migliore3} at this point, but we applied it to the DFT calculations which we will present in the next section. In contrast to the energy gap and effective coupling techniques, where $\varepsilon_{A}$=$\varepsilon_{D}$ has been assumed because the energies of the initial and final state need to be equal at the transition point where the transfer integral is defined in dependence on the reaction coordinate q, for the expansion coefficient method Eqn.~\ref{direct} has a discontinuity at this point, as is illustrated in the concrete example of Eqn.~\ref{adiabatic} which gives a$_{A,1}$=a$_{D,1}$=0.701. As it has been shown in the appendices of Refs.~\cite{migliore1} and~\cite{migliore2} this discontinuity can be eliminated leading to the expected correct result at the transition state coordinate but the closer the transition state is approached the higher the demands on the computational accuracy become. This leads to a trade off, where Eqn.~\ref{direct} is used close to but not at the transition point and for the model Hamiltonian in Eqn.~\ref{hamiltonian}, q can be varied by varying $\varepsilon_{D}$-$\varepsilon_{A}$.

\begin{table}[tp]
\begin{tabular}{|l|l|l|l|l||l|}
\hline \hline
$\varepsilon_{D}$ & $\varepsilon_{A}$ & $\varepsilon_{D}$-$\varepsilon_{A}$ & a$_{A,1}$ & a$_{D,1}$ & H$_{DA}^{coeff}$ \\
  \hline \hline
0.0 & 0.0 &  0.0 & 0.701 & 0.701 & divergent \\
-0.01  & 0.01 & 0.02 & 0.845 & 0.518 & 0.0197 \\
-0.1  & 0.1 & 0.2 & 0.991 & 0.093 & 0.0189 \\
-0.5  & 0.5 & 1.0 & 0.998 & 0.017 & 0.0167 \\
\hline \hline
\end{tabular}
\caption{Transfer integral H$_{DA}^{coeff}$ as calculated with the expansion coefficient method~\cite{migliore1}-~\cite{migliore3} in Eqn.~\ref{direct} for the Hamiltonian in Eqn.~\ref{hamiltonian} with the parameters $\varepsilon_B$=1, V$_{DB}$=V$_{BA}$=-0.1 and V$_{DA}$=-0.01. For the same parameters one can derive H$_{DA}^{gap}$=0.0195 from Eqn.~\ref{gap} and H$_{DA}^{effect}$=0.02 from Eqn.~\ref{effective} with $\varepsilon_A$=$\varepsilon_D$= 0.}
\label{tab.migliore}
\end{table}

It is illustrative to compare the values obtained for H$_{DA}$ from the three methods described in this section numerically for the Hamiltonian defined in Eqn.~\ref{hamiltonian} for the set of parameters which result in the adiabatic wavefunctions in Eqn.~\ref{adiabatic}. This is done in Table~\ref{tab.migliore}, where it can be seen that H$_{DA}^{coeff}$ converges towards 0.02 for small values of $\varepsilon_{D}$-$\varepsilon_{A}$, while H$_{DA}^{gap}$=0.0195 and H$_{DA}^{effect}$=0.02. As discussed in Ref.~\cite{voityuk1}, the applicability of the effective coupling method depends on $| \varepsilon_{B} - \varepsilon_{A,D} |$ being reasonably large and all couplings being reasonably small. In order to be more quantitative with this statement, we use the Hamiltonian in Eqn.~\ref{hamiltonian} with $\varepsilon_A$=$\varepsilon_D$=V$_{DA}$=0.0 and V$_{DB}$=V$_{BA}$ to derive H$_{DA}^{effect}=-V_{DB}^2 / \varepsilon_B$ and H$_{DA}^{gap}=0.5(0.5 \varepsilon_B - \sqrt{2 V_{DB}^2 + 0.25 \varepsilon_B^2}$ for this special case, which we both plot in dependence on V$_{DB}$ and $\varepsilon_B$ in Fig.~\ref{fig.gap_vs_eff}. It can be seen that the agreement between both methods is ideal for $| \varepsilon_{B} - \varepsilon_{A,D} |$ above 0.2 eV and $|V_{DB}|$ below 0.1 eV. Most systems we investigate in this article have Hamiltonians broadly within this range, but from Fig.~\ref{fig.gap_vs_eff} it can be also seen that the results from the two methods move away from each other only gradually for larger couplings or smaller onsite energy differences.

\begin{figure}
\includegraphics[width=0.9\linewidth,angle=0]{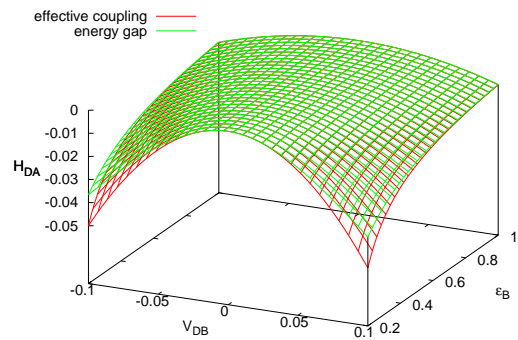}
\caption[cap.Pt6]{\label{fig.gap_vs_eff}H$_{DA}^{effect}$ and H$_{DA}^{gap}$ for the Hamiltonian in Eqn.~\ref{hamiltonian} with $\varepsilon_A$=$\varepsilon_D$=V$_{DA}$=0.0 and V$_{DB}$=V$_{BA}$.}
\end{figure}

\end{section}

\begin{section}{DFT calculations of the transfer integral for molecular benchmark systems}\label{sec:molecules}

\begin{figure}
\includegraphics[width=0.9\linewidth,angle=0]{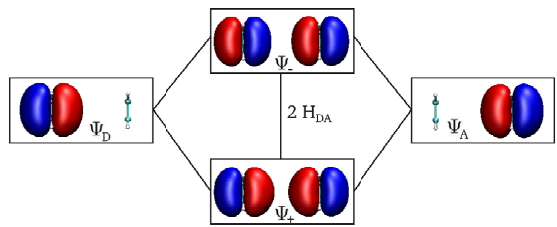}
\caption[cap.Pt6]{\label{fig.diethylene}Molecular orbitals for a dimer of ethylene molecules as studied in Ref.~\cite{bredas}, where the initial state $\psi_D$ for hole transport is the HOMO of the left molecule, and the final state $\psi_A$ the HOMO on the right one, and these orbitals form the adiabatic bonding and anti-bonding states, $\psi_+$ and $\psi_-$, respectively, through their hybridization.}
\end{figure}

\begin{figure}
\includegraphics[width=0.5\linewidth,angle=0]{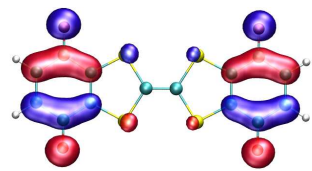}
\caption[cap.Pt6]{\label{fig.qttfq}The SOMO of Q-TTF-Q$^-$, which is the energetically lower lying adiabatic state for intra-molecular electron transfer in this system.}
\end{figure}

We now want to benchmark the three methods for evaluating H$_{DA}$, which we have introduced in the last section on real molecular systems instead of just TB matrices, where we choose two systems for which H$_{DA}$ has been studied extensively in the literature. The first is an ethylene dimer, where inter molecular hole transfer occurs between the local highest occupied MOs (HOMOs) of two ethylene molecules~\cite{bredas} which represent the diabatic initial and final state of the process, respectively, and form the two adiabatic bonding and antibonding states through their hybridization as we illustrate in Fig.~\ref{fig.diethylene}. The second test system is a Q-TTF-Q$^-$ anion, where intra molecular electron transfer between two quinone rings is mediated by a bridge~\cite{vanvoorhis} and the lower lying adiabatic state is shown in Fig.~\ref{fig.qttfq}. This latter case is more challenging to describe correctly, since i) it is not obvious whether the sulfur atoms should be seen as part of the initial/final state or as part of the bridge and ii) due to the direct covalent connection between the donor/acceptor states and the bridge states the self interaction (SI) problem of DFT, which results in an artificial tendency towards charge delocalization, becomes an issue~\cite{Perdew-Zunger,victor-paper,sanvito-victorpaper,graefenstein-victorpaper}. 

All DFT calculations in this article were performed with the GPAW code~\cite{GPAW1,GPAW2}, where the core electrons are described with the projector augmented wave (PAW) method and the basis set for the KS wavefunctions can be optionally chosen to be either a real space grid or a linear combination of atomic orbitals (LCAO), and we used both for the benchmarking calculations in the following, where the LCAO basis set has been applied on a double zeta level with polarisation functions (DZP). The sampling of the potential energy term in the Hamiltonian is always done on a real space grid when using GPAW, where we chose 0.18 \AA{} for its spacing and the same value when the grid also defines the basis set. For the XC functional we use the semi-local Perdew-Burke-Ernzerhof (PBE)~\cite{PBE} parametrisation but we compare it with the hybrid functional B3LYP~\cite{b3lyp} for the cases where we find an indication for an artificial delocalization of electronic states. A tool of GPAW we also use extensively in the following is the generalized $\Delta$SCF method, where the spatial expansion of an orbital enforced to contain a charge can be defined as an arbitrary linear combination of Bloch states~\cite{Schiotz1,Schiotz2}. By extracting or adding one electron from the system and inserting the corresponding charge into a predefined orbital in the beginning of every iteration step, the self consistency cycle progresses as usual but with the charge density of this particular orbital as a contribution to the external potential.

\begin{table}
\begin{center}
  \begin{tabular}{|c||c|c|c|c|c||c|}
\hline
basis set & \multicolumn{2}{c|}{H$_{DA}^{gap}$} & \multicolumn{1}{c|}{H$_{DA}^{effect}$} & \multicolumn{2}{c||}{H$_{DA}^{coeff}$} & H$_{DA}$ (Lit.) \\
\hline
&    \multicolumn{2}{c|}{SP \ MB} &  SP &   \multicolumn{2}{c||}{SP \ MB} &  \\
\hline
LCAO   &  0.030    & 0.026 & 0.033 &  0.033  & 0.043  & 0.046 \\
grid   &  0.050    & 0.043 &   -   &    -    & 0.068  & (Ref.~\cite{bredas}) \\
\hline
\end{tabular}
\end{center}
\caption{Transfer integral H$_{DA}$ for the ethylene dimer in Fig.~\ref{fig.diethylene} calculated with three different techniques, which are applied in single particle (SP) and many body (MB) variants, where the results are compared with those from Ref.~\cite{bredas} and are given in eV.}
\label{tab.diethylene}
 \end{table}

For the evaluation of the transfer integral with the energy gap method for the hole transport in the ethylene dimer one can in principle just obtain the two adiabatic states as the HOMO and HOMO-1 from a standard DFT calculation and insert their respective KS eigenenergies into Eqn.~\ref{gap}. We refer to this as a single particle (SP) approach in the following. Alternatively, one can use the $\Delta$SCF method in two separate calculations where an electron has been removed from either one or the other of these two orbitals in order to obtain total energies values whose insertion into Eqn.~\ref{gap} should ensure that H$_{DA}^{gap}$ calculated this way also contains contributions from the relaxation of all other electrons in reaction to this charge~\cite{CB1,CB2}. This is what we call the many body (MB) approach in the remainder of this article. In Table~\ref{tab.diethylene} we compare SP and MB values of H$_{DA}^{gap}$ for the ethylene dimer, where we calculated both with a LCAO as well as a grid basis set. 

For the definition of H$_{DA}^{effect}$ a subdiagonalization procedure~\cite{rectifier12,rectifier23} is required, where a Hamiltonian is obtained which contains a block with the onsite energies of orbitals localized on the left ethylene molecule and another block of states belonging to the right one with the couplings between left and right as non-diagonal elements. Since the ethylene dimer does not contain bridge states, only the direct coupling element between initial and final state, i.e. the first term on the right side of Eqn.~\ref{effective} is needed to obtain H$_{DA}^{effect}$ for this system. For this method we only define a SP mode, and consider MB calculations to be impractical.

While H$_{DA}^{gap}$ and H$_{DA}^{effect}$ have to be calculated at the ground state of the system with respect to the coordinates of the nuclei which corresponds to a reaction coordinate q=0, H$_{DA}^{coeff}$ diverges at this point as illustrated in the last section in the discussion of Table~\ref{tab.migliore}. In order to define a suitable q in terms of nuclear coordinates we followed the procedure in Ref.~\cite{vanvoorhis}, where the ground state coordinates ${\bf R}_0$ for the positively charged dimer are supplemented by relaxations for the charged initial and final states (with the charge localized through $\Delta$SCF on the left or right molecule, respectively) resulting in the sets of coordinates ${\bf R}_{-1}$ and ${\bf R}_{1}$ for q=-1 and q=1, respectively, and the interpolation formula ${\bf R}_q = 0.5 q (q+1) {\bf R}_1 - (q-1) (q+1) {\bf R}_0 + 0.5 q (q-1) {\bf R}_{-1}$ can be applied to obtain the coordinates for an arbitrary value of q. In the following we show results from calculations for q=0.2 wherever it is not stated otherwise. For calculating H$_{DA}^{coeff}$ in a SP mode we make use of the same block diagonalization of KS states already mentioned in connection with H$_{DA}^{effect}$ in the paragraph above, where the local HOMOs of the separate ethylene molecules in the dimer now have different energies due to the asymmetry of the system at q=0.2 and a finite energy difference can be obtained for Eqn.~\ref{direct}. By forming and diagonalizing a 2x2 Hamiltonian from these two local HOMOs and the direct coupling between them, the expansion coefficients a$_{A,1}$ and a$_{D,1}$ can also be straightforwardly derived. In MB mode $\epsilon_A$ and $\epsilon_D$ are the total energies of the initial and final state, respectively, and therefore $\Delta$SCF calculations constraining the positive charge on the left ethylene molecule at q=-1 and at the right one at q=1 have to be performed. The expansion coefficients a$_{A,1}$ and a$_{D,1}$ on the other hand are again quantities related to the transition point, and we use the wavefunction overlap within the projector augmented wave (PAW) formalism~\cite{GPAW2} to obtain them at q=0.2, where the coefficients of the expansion of adiabatic into diabatic states are equivalent to those for the expansion of constrained diabatic states into KS states if normalized correctly. We also tested the orthogonalisation procedures for the energy gap and the expansion coefficient methods suggested in Refs.~\cite{bredas} and ~\cite{migliore1}-~\cite{migliore3}, respectively, and found them to have no numerical effect for any system studied in this article, where all states in the definitions we chose were orthogonal already.

\begin{table}
\begin{center}
  \begin{tabular}{|c||c|c|c|c|}
\hline
&    H$_{DA}^{gap}$ (MO) & H$_{DA}^{gap}$ (diag.) & H$_{DA}^{effect}$ & H$_{DA}^{coeff}$ \\
\hline
S on donor/acceptor &  0.031 & 0.023 & 0.023 & 0.023 \\
S on bridge          &  0.031 & 0.042 & 0.064 & 0.057 \\
\hline
\end{tabular}
\end{center}
\caption{Transfer integral H$_{DA}$ for the Q-TTF-Q$^-$ anion in Fig.~\ref{fig.qttfq} calculated with all three techniques in SP mode, where the Sulfur atoms are taken as part of the initial and final state in the first row, and as part of the bridge states in the second row. All numbers in this table have been calculated with a LCAO basis set and are given in eV.}
\label{tab.qttfq_single}
 \end{table}

From Table~\ref{tab.diethylene} it can be seen that for the ethylene dimer all three methods agree perfectly with each other in SP mode, where only the energy gap technique can be applied also with a grid basis set, while for the other two approaches the LCAO basis is needed for the subdiagonalization procedure of KS states~\cite{rectifier12,rectifier23}. There is a bit more fluctuation of results in MB mode but overall the deviations are moderate, where more accuracy tends to deliver slightly higher values assuming that the grid basis is better converged than the LCAO basis and that MB in general gives an improvement over SP. We also show the number obtained by Bredas and co-workers for the same system in Ref.~\cite{bredas}, which matches perfectly with our MB values of H$_{DA}^{coeff}$ with a LCAO and H$_{DA}^{gap}$ with a grid basis. One might wonder why the MB values do not differ more when compared with their SP counterparts given that electronic relaxation provides a factor of two when e.g. comparing the addition energy of a H$_2$ molecule in a junction with the molecules KS-HOMO/LUMO gap~\cite{CB1}. This discrepancy is best understood by focusing on Eqn.~\ref{gap} for the calculation of H$_{DA}^{gap}$ from two separate total energy calculations with a positive charge in first the energetically lower and than the higher of the two adiabatic states in Fig.~\ref{fig.diethylene} which are the global HOMO and HOMO-1 of the dimer. Contrarily to the bonding HOMO and anti-bonding LUMO of H$_2$ which differ considerably in their respective spatial distribution, the HOMO and HOMO-1 of the ethylene dimer mostly differ in their phase, i.e their minima and maxima are at exchanged positions for the second ethylene molecule. This is, however, irrelevant for the electron density that is formed from these orbitals where the minima and maxima both give local peaks and the effect of the relaxation of the other electrons in the system should be similar for both states, thereby almost cancelling out when the difference in Eqn.~\ref{gap} is formed. In general such a good agreement between the SP and MB mode of the energy gap method can always be expected because transfer integrals are usually below 0.1 eV in value, which corresponds to a rather low level of hybridization between the donor and acceptor states and therefore to rather similar spatial distributions of the bonding and antibonding adiabatic states.

It has to be stressed that the ethylene dimer is a rather unchallenging system in the sense that the initial and final state are not connected to each other by covalent bonds and therefore no bridge states exist. The difficulties that can arise in the presence of a bridge are illustrated for the Q-TTF-Q$^-$ anion in Table~\ref{tab.qttfq_single}, where H$_{DA}$ has been calculated in SP mode with all three methods. Although like in the dimer case, also for the anion the energy difference of the SOMO and LUMO from the standard DFT calculation can be used directly for determining H$_{DA}^{gap}$ for electron transfer, there is an ambiguity here because the initial and the final state have not been explicitly defined and in principle several diabatic states localized on the two quinone rings could contribute to what we take as the adiabatic states. This ambiguity can be overcome by block-diagonalizing the KS Hamiltonian over the donor, bridge and acceptor areas, selecting one state in the donor area as initial and one in the acceptor area as final state, keeping all N bridge states, and then diagonalizing the resulting (N+2)x(N+2) Hamiltonian, where the two adiabatic states for forming the energy difference can be chosen by the criterion of a high amplitude at the initial and final state as discussed in the previous section. We distinguish between these two ways of deriving H$_{DA}^{gap}$ in SP mode just described by referring to them as H$_{DA}^{gap}$ (MO) and H$_{DA}^{gap}$ (diag.). The latter becomes especially important in the next section, where we have the Bloch states of the gold leads as initial and final states and their selection becomes a crucial issue for the transfer integral. The same (N+2)x(N+2) Hamiltonian is also relevant for the derivation of H$_{DA}^{effect}$, where now the first term in Eqn.~\ref{effective} gives only a negligible contribution and the N bridge states are all entering into the sum. Also H$_{DA}^{coeff}$ we obtain by diagonalizing a (N+2)x(N+2) Hamiltonian for the expansion coefficients but this one now represents the electronic structure for the nuclear coordinates corresponding to q=0.2. 

The most important question if a molecular bridge exists between the donor and acceptor states is the decision which atoms are still part of the initial/final state and which atoms should be assigned to the bridge. Although this decision is in principle arbitrary if all parts of the system are connected by covalent bonds, for some systems there are natural choices as we discuss in the next section where the initial and final state are on the gold leads and the molecule is the bridge. For the Q-TTF-Q$^-$ anion there is no a priori way to make a superior assignment for the sulfur atoms and we compare the results for both possibilities in Table~\ref{tab.qttfq_single}. For H$_{DA}^{gap}$ (MO) there is no difference because we do not describe our initial and final states explicitly as stated above and therefore do not specify where their location starts and ends. For all other methods the values for the transfer integral vary between the two choices of what is the bridge as they should. The correct value of H$_{DA}$ has to depend on the exact definition of A and D or in other words the transfer integral for two quinone rings with sulfur ends connected by an ethylene bridge is different from the one for two quinone rings connected by an ethylene tetrathiol bridge. More interestingly, while all three methods give the same result with the S atoms as part of donor and acceptor, they exhibit quite a spread of results if these atoms are part of the bridge. This finding can be explained by coming back to the discussion around Fig.~\ref{fig.gap_vs_eff} where it has been shown that the methods only give the same results for reasonably small couplings and reasonably large energy differences. If the S atoms are considered to be part of the bridge, the couplings reach values of up to 0.8 eV and therefore the methods slightly diverge for this case.

\begin{table}
\begin{center}
  \begin{tabular}{|c|c|c||c|}
\hline
&    H$_{DA}^{gap}$ & H$_{DA}^{coeff}$ & H$_{DA}$ (Lit.) \\
\hline
PBE &  0.026 & 0.117 (0.157) & 0.130 \\
B3LYP & 0.036 & 0.053 (0.035) & (Ref.~\cite{vanvoorhis}) \\
\hline
\end{tabular}
\end{center}
\caption{Transfer integral for the Q-TTF-Q$^-$ anion (at q=0 and q=0.2 for H$_{DA}^{gap}$ and H$_{DA}^{coeff}$, respectively) calculated with PBE and B3LYP functionals in MB mode. All results in this table have been obtained with a grid basis set and are given in eV. The result of Ref.~\cite{vanvoorhis} for this system is also shown for comparison.}
\label{tab.qttfq_XC}
 \end{table}

In Table~\ref{tab.qttfq_XC} we show H$_{DA}$ for the Q-TTF-Q$^-$ anion calculated with the energy gap and expansion coefficient techniques in MB mode. The main numbers for H$_{DA}^{coeff}$ have been calculated with the sulfur atoms as part of the initial and final state, while a definition with the S atoms being part of the bridge has been used for the numbers in parantheses. All results we presented so far have been produced with a PBE~\cite{PBE} parametrisation of the XC functional, while in Table~\ref{tab.qttfq_XC} we also compare with data using the hybrid functional B3LYP~\cite{b3lyp} instead. It can be seen that the PBE version of H$_{DA}^{coeff}$ deviates from all the other values we have calculated for the transfer integral in the Q-TTF-Q$^-$ anion by an order of magnitude but interestingly is almost equal to the value found in Ref.~\cite{vanvoorhis}. The explanation of this deviation can be found in the SI problem, which makes the expansion coefficients a$_{D,1}$ and a$_{A,1}$ almost equal even if asymmetry is induced by setting q=0.2. This artefact can be even more highlighted by calculating the expansion coefficients at q=-1, where one of them should be close to 0 and the other one close to 1, which is indeed the case for B3LYP (a$_{D,1}$=0.95, a$_{A,1}$=0.001) but not for PBE (a$_{D,1}$=0.77, a$_{A,1}$=0.64). This problem does not occur for the PBE calculations of H$_{DA}^{coeff}$ for the ethylene dimer presented in Table~\ref{tab.diethylene} where a$_{D,1}$=0.99 and a$_{A,1}$=0.10, because in this case there is no bridge linking the donor and acceptor~\cite{victor}. Since in the expansion method the diabatic states are defined as a linear combination of the adiabatic states, the SI error cannot lead to an artificial overdelocalisation where the charges are already maximally delocalized over donor and acceptor (as is the case for the ethylene dimer) but it has an effect on the Q-TTF-Q$^-$ anion where the charge can spill onto the bridge. 

In summary, we can conclude from this section that all three methods agree with each other quite well for the chosen benchmark systems and that results from single particle and many body calculations are of the same order of magnitude. We therefore restrict our study to the SP mode in the remainder of this article.

\end{section}

\begin{section}{Calculation of the transfer integral for a redox active single molecule junction}\label{sec:junction}

\begin{figure}
      \includegraphics[width=0.95\linewidth,angle=0]{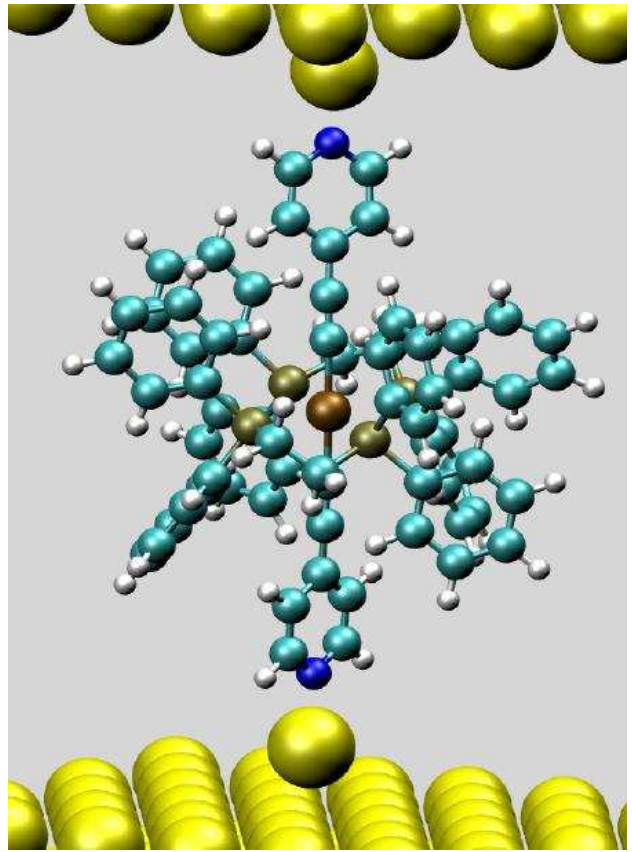}
        \caption[cap.Pt6]{\label{fig.complex}Geometry of the Ru(PPh$_2$)$_4$(C$_2$H$_4$)$_2$ bis(pyridylacetylyde) complex bonded to ad-atoms on Au fcc (111) surfaces, where the conductance has been studied within the framework of NEGF for the neutral and oxidized complex in Ref.~\cite{first}.}
\end{figure}

\begin{table*}
 \begin{center}
\begin{tabular}{|c||c|c|c||c|c|c|c||c|c|c|c|}
\hline
Charge & \multicolumn{3}{c||}{Au pyramid} & \multicolumn{4}{c||}{${\bf{\Gamma}}$ point only} & \multicolumn{4}{c|}{8 {\bf k} points}  \\
\hline
& H$_{DA}^{gap}$ & H$_{DA}^{effect}$ & H$_{DA}^{coeff}$ & H$_{DA}^{gap}$& H$_{DA}^{effect}$ & H$_{DA}^{coeff}$ & H$_{DA}^G$ & H$_{DA}^{gap}$ &H$_{DA}^{effect}$ & H$_{DA}^{coeff}$ &H$_{DA}^G$ \\
\hline
0& 3.57 10$^{-3}$ & 3.63 10$^{-3}$ & 3.65 10$^{-3}$ & 2.95 10$^{-4}$ & 3.05 10$^{-4}$ & 3.05 10$^{-4}$ & 4.41 10$^{-4}$ & 1.06 10$^{-3}$ &1.02 10$^{-3}$ & 1.02 10$^{-3}$ & 4.96 10$^{-4}$ \\
+1 & 3.16 10$^{-3}$ & 3.26 10$^{-3}$& 3.18 10$^{-3}$ & 4.28 10$^{-4}$ & 4.72 10$^{-4}$ & 4.69 10$^{-4}$ & 1.00 10$^{-3}$ & 1.95 10$^{-3}$ & 1.65 10$^{-3}$ & 1.25 10$^{-3}$ & 1.37 10$^{-3}$ \\
\hline
  \end{tabular}
\caption{Transfer integrals for the junction in Fig.~\ref{fig.complex} calculated with energy gap, effective coupling and expansion coefficient methods in SP mode. The gold leads are small clusters of four atoms on each side in the Au pyramid columns and the periodic surfaces from Ref.~\cite{first} everywhere else, where ${\bf{\Gamma}}$ point only calculations are also compared with the average over 8 {\bf k} points in the irreducible Brillouin zone. $H_{DA}^G$ has been defined as $\sqrt{G(E_F)}/8$, where the values of G(E$_F$) have been taken from Ref.~\cite{first}. All results are given in eV.}
\label{tab.alldata}
 \end{center}
 \end{table*} 

\begin{figure}
      \includegraphics[width=0.95\linewidth,angle=0]{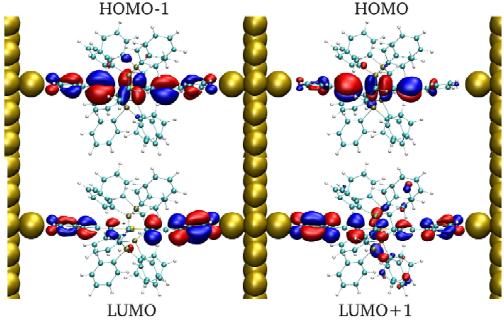}
        \caption[cap.Pt6]{\label{fig.mo}Molecular orbitals close to E$_F$ for the junction in Fig.~\ref{fig.complex}.}
\end{figure}

\begin{table}
\begin{center}
\begin{tabular}{|c||c|c|c|c|c|}
\hline
Charge  &    HOMO-1        &    HOMO        & LUMO        &    LUMO+1        \\
\hline
0        &    3.4x10$^{-4}$    &    3.3x10$^{-5}$ &    1.1x10$^{-3}$    &    -5.8x10$^{-4}$  \\
\hline
+1        &    1.4x10$^{-3}$   &    4.4x10$^{-5}$ &    6.2x10$^{-4}$    &    -4.7x10$^{-4}$  \\
\hline
\end{tabular}
\caption{Individual contributions of the MOs in Fig.~\ref{fig.mo} to H$_{DA}^{effect}$. All values are given in eV.}
\label{tab.mo}
\end{center}
\end{table} 

In a recent article we studied the coherent electron transport through the Ru(PPh$_2$)$_4$(C$_2$H$_4$)$_2$ bis(pyridylacetylyde) complex in Fig.~\ref{fig.complex} by using a NEGF formalism for the conductance, and where we switched the oxidation state of the redox active ruthenium atom between +II and +III corresponding to an overall charge of 0 and +1 on the molecular complex, respectively~\cite{first}. The +1 charge on the complex we achieved by fixing the charge on a Cl counter ion with the $\Delta$SCF technique~\cite{Schiotz1,Schiotz2} and ensuring overall charge neutrality in the unit cell of our device region, so that the negative charge on the chlorine anion resulted in a compensating positive charge on the complex. In Ref.~\cite{first} we also tested a second method for charge localisation, which made use of solvent screening and was computationally more expensive but in the current article no water molecules were added to our cell, because the more efficient approach based on $\Delta$SCF achieved equivalent or even better results. We chose this particular Ru-complex because it was used in previous conductance measurements~\cite{Frisbie, Wang} as a monomer of chains, where depending on the chain length either coherent transport or electron hopping was observed~\cite{Frisbie}, and because this molecular species is in general considered to be a good starting point for the investigation of chains with multiple redox active centers~\cite{Dixneuf}. In contrast to Ref.~\cite{Frisbie} we use pyridil groups as anchors to the leads because they provide peaks in the transmission function, which are narrow enough to assume that a charge on the complex has an impact on the conductance but broad enough to avoid the Coulomb blockade regime~\cite{robert-21,robert-20,robert-6}.

In the present article we relate the conductance G of the molecular junction in Fig.~\ref{fig.complex} to the transfer integral, where the electrodes play the role of the initial and the final state in a one-step electron transfer reaction and the redox-active molecule acts as a mediating bridge. The relation between H$_{DA}$ and G was explicitly described by Nitzan~\cite{nitzan1,nitzan2}, where adopted to our definition of H$_{DA}$ the conductance was expressed as
\begin{equation}
 G(E_F) = \frac{H_{DA}^2 \Gamma_A \Gamma_D G_0}{((E_F-E_D)^2+\Gamma_D^2/4)((E_F-E_A)^2+\Gamma_A^2/4)},
\label{nitzan_eq}
\end{equation}
with G$_0$ being the conductance quantum, and $\Gamma_D$ and $\Gamma_A$ the widths of the donor and acceptor levels due to their couplings to the left and right metal leads, respectively. In such a setup only metallic surface states close to $E_F$ are relevant for the conductance through the junction and if only such bands are considered as the initial and final state of the corresponding electron transfer reaction the energy differences $E_F-E_D$ and $E_F-E_A$ vanish in Eqn.~\ref{nitzan_eq}, which can now be simplified to
\begin{equation}
 G(E_F) = H_{DA}^2 \frac{16}{\Gamma_D \Gamma_A} G_0.
\label{nitzan_eq2}
\end{equation}
By using this expression and setting $\Gamma_D=\Gamma_A=0.5 eV$, which is reasonable for the coupling width of gold leads~\cite{troels}, the transfer integral can be obtained from the conductance as $H_{DA}^G \approx \sqrt{G(E_F)}/8$, where in Table~\ref{tab.alldata} we list the values we derive in this way from the conductances in Ref.~\cite{first} as a benchmark for the three methods introduced in the current article.

In order for Eqn.~\ref{nitzan_eq2} to be valid only metal bands which contribute to the density of states (DOS) at the Fermi Level can be considered as donor and acceptor states. Therefore, we calculated H$_{DA,\mathbf{k}}$ for all relevant donor-acceptor pairs weighted with a {\bf{k}}-point resolved DOS,
\begin{equation}
 H_{DA,\mathbf{k}} = \frac{\sum_{i=1}^N H_{DA,i,\mathbf{k}}*\rho(E_F)_{i,\mathbf{k}}}{\sum_{i=1}^N \rho(E_F)_{i,\mathbf{k}}, },
\label{dos_mean}
\end{equation}
where $\rho$ is the density and finally integrated H$_{DA,\mathbf{k}}$ over {\bf{k}}-space following the procedure of Marcus and co-workers~\cite{kpoints}. Another aspect which has to be considered for the proper choice of initial and final states is their localization on the gold adatoms which couple directly to the molecular bridge, since only those can contribute signficantly to the electron transfer reaction. Bulk bands, which are close to the Fermi Level but have no connection to the molecule would lower H$_{DA}$ in Eqn.~\ref{dos_mean} artificially, where the statistical weight is only defined by the DOS. Therefore we introduced the exclusion criterion that the metallic states entering Eqn.~\ref{dos_mean} have a coupling with one of the two most relevant MOs (the HOMO-1 and the LUMO), which is larger than 10$^{-3}$. 

In theory, the initial and the final state in our calculations should have the same energy for each donor/acceptor pair, because the junction in Fig.~\ref{fig.complex} has a high symmetry and the transfer integral has to be defined at the transition point of the corresponding reaction. In practice, however, small asymmetries introduced by the torsional degrees of freedom in the molecular bridge can lead to differences in diabatic energies in the range of 10$^{-3}$ eV. Since the H$_{DA}$ values in this section are in the order of 10$^{-4}$ eV, we corrected Eqn.~\ref{gap} to account for these asymmetries following the procedure of Bredas and co-workers~\cite{bredas}, where the differences of the diabatic energies are explicitly subtracted,
\begin{equation}
 H_{DA}^{gap}= \frac{\sqrt{(\varepsilon_1-\varepsilon_2)^2 - (\epsilon_A-\epsilon_D)^2}}{2}.
\label{gap2}
\end{equation}
For the application of the expansion coefficient method, it is an advantage that small energy differences between donor and acceptor states exist, because here we can interpret them as finite values of q, since this technique cannot be applied at the transition point as discussed in detail in the previous two sections. The effective coupling method can be corrected by replacing the denominator $\varepsilon_{A,D}-\varepsilon_i$ of the second term in Eqn.~\ref{effective} with $(\varepsilon_{D}+\varepsilon_{A})/2-\varepsilon_i$~\cite{voityuk1}.
Another consequence of the asymmetry in the junction are artificial deviations between the couplings of each MO to the two gold surfaces, which we corrected for by using a mean value for the coupling to both surfaces for all three techniques.

In Table~\ref{tab.alldata} we present H$_{DA}$ values for the junction in Fig.~\ref{fig.complex} for a neutral (0) and charged (+1) complex calculated with all three methods at only the ${\bf \Gamma}$ point as well as averaged over the 8 {\bf k} points in the irreducible Brillouin zone obtained from a 4x4x1 grid. The values H$_{DA}^G$ are obtained from the conductances in Ref.~\cite{first} for the same system and also given for comparison. While all the remaining numbers refer to the periodic junction also used in Ref.~\cite{first}, for the first three columns (Au pyramid) small clusters of 4 gold atoms on each side in a tetraeder configuration have been used as electrodes in order to assess the effect of a proper description of the gold leads on the numbers. The overall agreement of the three methods for calculating H$_{DA}$ amongst themselves is excellent, which is not surprising because all relevant couplings between the MOs of the bridge and the surface states are in the range 10$^{-3}$-10$^{-2}$ and the molecular eigenenergies have at least a distance of 0.2 eV from E$_F$. When compared with $H_{DA}^G$ the important aims are fulfilled, namely the order of magnitude is the same, and the transfer integral for the neutral state is always considerably smaller than that for the charged one. A better agreement could not have been expected given that the values for $H_{DA}$ are rather small and the approximative nature of the assumptions we made in deriving H$_{DA}^G$ from G (E$_F$). This is in particular true for the underestimation of the {\bf k} point dependence of H$_{DA}^G$ in Table~\ref{tab.alldata}, which stems from the fact that $\Gamma_D$ and $\Gamma_A$ in Eqn.~\ref{nitzan_eq2} depend on the density of states of the lead and should therefore be different for each {\bf k} point, which is not considered in our treatment, where we set both to 0.5 eV globally throughout the reciprocal space of the system. Only the results we obtained for the transfer integral with small clusters as gold leads are wrong, both in their order of magnitude and in the ranking with regard to 0 and +1 charge, where both can be easily explained. The small cluster size is responsible for a larger amplitude of the initial/final state on the Au ad-atom, thereby enhancing the coupling to all MOs~\cite{troels} and resulting in artificially high values of H$_{DA}$. The charged complex does not have higher transfer integrals then the neutral one because for non-periodic leads the charge introduced by the chlorine ion is mostly localized on the Au clusters as we investigated in detail in Ref.~\cite{first}.

An important finding from the comparison of conductances of the two charging states of the Ru-complex in Ref.~\cite{first} was that for the neutral molecule it was determined by the molecular LUMO and LUMO+1, with the contribution from the LUMO being distinctly larger. In the charged case the molecular HOMO and HOMO-1 are shifted close to the Fermi energy of the metal leads, which makes them primarily responsible for the molecular conductance. The effective coupling method provides a good way to analyze whether the same holds true for the respective transfer integrals because the contributions from the MOs are additive in Eqn.~\ref{effective}. In Table~\ref{tab.mo} we list the terms in the sum coming from the four MOs closest to E$_F$, where indeed it can be seen that the LUMO dominates for charge 0 and the HOMO-1 for charge +1. The HOMO adds only an amount which is two orders of magnitude smaller for both oxidation states, because it is mostly localized in the center of the molecule and only to a much lesser extent on the anchor groups as can be seen from Fig.~\ref{fig.mo}. This results in rather low coupling of this orbital to the metal leads, which we also found from the NEGF calculations for the transmission functions in Ref.~\cite{first}. The contribution from the LUMO+1 is also independent of the charging state but of larger magnitude than that of the HOMO, and for both systems its sign is different from that of the other MOs indicating destructive interference. 

\end{section}

\begin{section}{Summary}\label{sec:summary}

The aim of this article was to identify suitable methods for calculating the transfer integral -which is a crucial quantity in Marcus theory- within a DFT framework for a setup where metallic leads act as donor and acceptor and a molecule in between them mediates electron or hole transport as a bridge. We found three techniques fit for that purpose, namely i) the energy gap method where H$_{DA}$ is derived from the total energy difference of adiabatic states, ii) Larsson's formula which adds up the contributions from each MO of the molecular bridge and iii) a expansion coefficient approach where the amplitudes of the adiabatic states in a diabatic basis are used explicitly. For this assessment we proceeded in three steps. First we compared the three methods on an abstract level by applying them to 3x3 tight binding matrices, where we found good agreement between all of them for small couplings between the bridge and the donor/acceptor states and large respective onsite energy differences. In a second step we benchmarked our DFT implementation of the three techniques for purely molecular systems with and without a bridge which have been studied by other groups, where we also established that a many body approach gives only negligible corrections compared to single particle descriptions. Finally, we calculated H$_{DA}$ for a single molecule junction where a Ru complex is coupled to two gold surfaces by pyridyl anchor groups using all three methods and assuming that surface states of the two leads act as donor and acceptor states, thereby describing coherent tunnelling. Our results for H$_{DA}$ were in excellent agreement with those derived from the conductance computed with a NEGF formalism for the same system in two different oxidation states.

\end{section}


\begin{acknowledgments}
G.K. and R.S. are currently supported by the Austrian Science Fund FWF, project Nr. P22548. We are deeply indebted to the Vienna Scientific Cluster VSC, on whose computing facilities all calculations presented in this article have been performed (project Nr. 70174) and where we were provided with extensive installation and mathematical library support by Markus St\"{o}hr and Jan Zabloudil in particular. We are deeply indebted to Pawel Zawadzki for sharing his expert knowledge in applying the generalized $\Delta$SCF technique as implemented in the GPAW code with us, and gratefully acknowledge helpful discussions with Tim Albrecht.
\end{acknowledgments}


\bibliographystyle{apsrev}

\end{document}